# Cross-cultural electronic word-of-mouth: a systematic literature review

Poompak Kusawat and Surat Teerakapibal

*Thammasat University, Bangkok, Thailand*

## Abstract

**Purpose** – Global adoption of the internet and mobile usage results in a huge variation in the cultural backgrounds of consumers who generate and consume electronic word-of-mouth (eWOM). Unsurprisingly, a research trend on cross-cultural eWOM has emerged. However, there has not been an attempt to synthesize this research topic. This paper aims to bridge this gap.

**Methodology** – This research paper conducts a systematic literature review of the current research findings on cross-cultural eWOM. Journal articles published from 2006 to 2021 are included. This study then presents the key issues in the extant literature and suggests potential future research.

**Findings** – The findings show that there has been an upward trend in the number of publications on cross-cultural eWOM since the early 2010s, with a relatively steeper increase toward 2020. The findings also synthesize cross-cultural eWOM research into four elements and suggest potential future research avenues.

**Value** – To the best of the authors' knowledge, there is currently no exhaustive/integrated review of cross-cultural eWOM research. This research fills the need to summarize the current state of cross-cultural eWOM literature and identifies research questions to be addressed in the future.

**Paper type** Literature review

## Resumen

**El boca a boca electrónico cross-cultural: una revisión sistemática de la literatura**

**Objetivo** – La adopción global de Internet y los móviles da lugar a una enorme diferencia en el origen cultural de los consumidores que generan y consumen el boca a boca electrónico (eWOM). No es de extrañar que haya surgido una tendencia de investigación sobre el eWOM transcultural. Sin embargo, no se ha intentado sintetizar este tema de investigación. El objetivo de este artículo es subsanar esta carencia.

**Metodología** – Este trabajo de investigación realiza una revisión bibliográfica sistemática de las investigaciones realizadas sobre eWOM transcultural. Se incluyen artículos de revistas publicados desde 2006 hasta 2021. A continuación, el estudio presenta las cuestiones clave de la literatura existente y sugiere posibles investigaciones futuras.

**Resultados** – Los resultados muestran que ha habido una tendencia al alza en el número de publicaciones sobre eWOM intercultural desde principios de la década de 2010, con un aumento relativamente creciente hacia 2020. Los resultados también sintetizan la investigación sobre eWOM intercultural en cuatro elementos y sugieren posibles vías de investigación futuras.

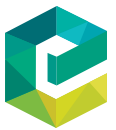

**Valor** – Actualmente no existe una revisión exhaustiva/integrada de la investigación sobre el eWOM cross-cultural. Esta investigación satisface la necesidad de resumir el estado actual de la literatura sobre eWOM cross-cultural e identifica las cuestiones de investigación que deben abordarse en el futuro.

**Palabras clave** – Revisión de la literatura, Boca a boca electrónico, Revisión online, Contenido generado por el usuario, Diferencias culturales, Cros-cultural

**Tipo de artículo** – Revisión de la literatura

跨文化电子口碑研究：系统性文献回顾

摘要

目的 – 在互联网全球化以及移动手机的广泛使用的背景下，不同文化背景的消费者都在贡献电子口碑（eWOM）。这使得电子口碑存在文化差异。然而，还没有人试图对这个研究课题进行综合分析。本文的目的就是要弥补这一空白。

方法 – 本研究论文对目前关于跨文化eWOM的研究成果进行了系统的文献回顾。包括2006年至2021年发表的期刊文章。然后，本研究提出了现有文献中的关键问题，并提出了潜在的未来研究。

研究结果 – 研究结果显示，自2010年初以来，关于跨文化eWOM的出版物数量呈上升趋势，到2020年时增幅相对较大。研究结果还总结了跨文化eWOM研究的四个要素，并提出了潜在的未来研究途径。

价值 – 目前还没有关于跨文化eWOM研究的详尽/综合的回顾。这项研究填补了总结跨文化电子WOM文献现状的需要，并确定了未来要解决的研究问题。

关键词 文献综述，电子口碑，在线评论，用户生成的内容，文化差异，跨文化

文章类型：研究型论文

## Introduction

With the rise of the internet, consumer behaviors regarding the search for information and interaction with others have changed drastically. As a result, a new research stream called electronic word-of-mouth (eWOM) has emerged. Initially, when consumers seek product information, they often turn to firm-generated information or reach out to friends and families for advice. The emergence of eWOM has allowed potential, actual and former customers to make positive or negative statements about a product or company to a multitude of people and institutes via the internet (Hennig-Thurau *et al.*, 2004). Consumers across the globe are increasingly using the internet and participating in eWOM. By April 2021, the number of internet users had grown by 7.6% from the previous year to reach 4.72 billion. This number equates to more than 60% of the world's total population (Kemp, 2021). This global adoption of the internet has resulted in a huge variety of cultural backgrounds of consumers who write and/or read eWOM.

This phenomenon is highly relevant to marketers. While it allows marketers to reach customers to an extent never seen before, what works well in one country could be disastrous in another. Marketing attempts with too little awareness of cross-cultural differences can result in misunderstandings, hurt feelings and communication errors that could lead to serious damage (Tone *et al.*, 2009). To better understand this phenomenon, the number of publications focusing on cross-cultural eWOM has been rising. However, current research on cross-cultural eWOM is rather fragmented. In addition, researchers have adopted various cultural theories in their studies, making it difficult to draw meaningful conclusions. This is detrimental to the systematic development of knowledge and the consolidation of the extant research findings.

To bridge this gap, this study offers a systematic review of the extant empirical research to examine how the relationship between cultural values and eWOM has been discussed in the literature and summarizes all available empirical evidence on the reviewed relationships.

To achieve this, we propose two key research questions. Answering these questions can help the reader understand all relevant information and identify which culture–eWOM relationship patterns are strongly and consistently supported by the available evidence. Additionally, we point out where such evidence is inconclusive, where meaningful evidence may be completely absent and where further research is needed. The research questions of this study are stated below:

*RQ1*. How do the cultural backgrounds of eWOM communicators affect the content of and responses to eWOM?

*RQ2*. How do the cultural backgrounds of eWOM receivers affect their responses to eWOM?

By addressing these research questions, we contribute to both academia and practitioners. To our knowledge, this is the first study to provide a comprehensive review of the existing eWOM literature in the cross-cultural context. Though there have been various reviews of eWOM literature (for example, Cheung and Thadani, 2012; Verma and Yadav, 2021), these studies rarely, if at all, focus on the cultural aspects of eWOM. By conducting a thorough review, we are able to classify a wide range of culture–eWOM relationships. This provides interested researchers with up-to-date knowledge on cross-cultural eWOM and can serve as an important foundation for future research. Moreover, this paper extends Cheung and Thadani's (2012) framework by including a cultural aspect, thereby facilitating the development of the body of knowledge on eWOM. For global marketers, our paper provides a guideline of how consumers communicate and/or respond to eWOM across cultures. As communication channels have shifted from offline to online, comprehending the characteristics of cross-cultural eWOM is the first step to harness the power of the Web 2.0 era.

This paper is organized into five sections. The next section discusses the background of cross-cultural eWOM studies. The third section describes our research. Section four presents the results. Finally, section five discusses the implications, limitations and future research.

## Background
Word-of-mouth (WOM) is one of the oldest ways people spread information. The emergence of online platforms gave rise to a new form of communication, eWOM. Though various definitions of eWOM exist in the literature, in this study, we define eWOM as "any positive or negative statement made by potential, actual or former customers about a product or company, which is made available to a multitude of people and institutions via the internet" (Hennig-Thurau *et al.*, 2004, p. 39). Although many consider eWOM to be the electronic version of traditional WOM, certain differences between the two concepts exist. First, eWOM communications can reach a larger number of consumers in a shorter period of time than WOM communications can. Second, eWOM is persistent and stays in public repositories, meaning it is always available to potential consumers who are actively searching for information about products or services. Third, eWOM is more balanced and unbiased than traditional WOM because various opinions are displayed at the same time on the same platform. Finally, traditional WOM receivers generally know the identity of the senders. Therefore, the credibility of the communicator and the message is known to the receiver.

It is widely recognized that culture is an important factor affecting consumers' thoughts and actions (De Mooij and Hofstede, 2011). Unsurprisingly, a stream of published studies has recently emerged investigating the influence that cultural differences have on eWOM. The pioneering cross-cultural eWOM research was conducted by Fong and Burton (2006). By collecting data from American and Chinese product discussion boards, they attempted to understand information-seeking behavior across cultures. Since then, a variety of cultural theories have been adopted in the research, for instance, Hofstede's cultural dimensions theory (Hofstede, 1980; Hofstede *et al.*, 2010), the global leadership and organizational behavior effectiveness (GLOBE) cultural framework (House *et al.*, 2004), Schwartz's cultural values (Schwartz, 1992), horizontal and vertical dimensions of individualism and collectivism (HVIC) (Singelis *et al.*, 1995; Triandis and Gelfand, 1998), high/low cultural context (Hall, 1976) and analytical/holistic thinking style (Nisbett *et al.*, 2001). Among them, Hofstede's is the most widely used. Definitions of cultural theories used in prior cross-cultural eWOM research are as follows:

Hofstede's cultural dimensions (Hofstede, 1980; Hofstede *et al.*, 2010):

- Power distance – the degree to which the less powerful accept that power is distributed unequally.
- Individualism – the degree of interdependence a society maintains among its members.
- Masculinity – the degree of preference in achievement, heroism, assertiveness and material rewards for success.
- Uncertainty avoidance – the degree to which a society feels uncomfortable with uncertainty and ambiguity.
- Long term-orientation – the fostering of virtues oriented towards future rewards.
- Indulgence – a society that allows natural human drives related to enjoying life and having fun.

GLOBE (House *et al.*, 2004):

- Performance orientation – the degree to which a collective encourages and rewards group members for performance.
- Assertiveness – the degree to which individuals are assertive and aggressive in their relationship with others.
- Future orientation – the extent to which individuals engage in future-oriented behaviors.
- Humane orientation – the degree to which a collective encourages and rewards individuals for being fair and kind to others.
- Institutional collectivism – the degree to which organizational and societal institutional practices encourage and reward collective distribution of resources and collective action.
- In-group collectivism – the degree to which individuals express pride, loyalty and cohesiveness in their organizations.
- Gender egalitarianism – the degree to which a collective minimizes gender inequality.
- Power distance – the extent to which the community accepts status privileges.
- Uncertainty avoidance – the extent to which a society relies on social norms and rules to alleviate unpredictability of future events.

Schwartz (1992):

- Openness to change vs conservation – the conflict between independence (self-direction, stimulation) and obedience (security, conformity, tradition).
- Self-transcendence vs self-enhancement – the conflict between the concern for the welfare of others (universalism, benevolence) and one's own interests (power, achievement).

HVIC (Triandis and Gelfand, 1998; Singelis *et al.*, 1995):

- Horizontal individualism – individuals strive to be distinct without desiring special status.
- Horizontal collectivism – individuals emphasize interdependence but do not submit easily to authority.
- Vertical individualism – individuals strive to be distinct and desire special status.
- Vertical collectivism – individuals emphasize interdependence and competition with out-groups.

Cultural context (Hall, 1976):

- A high-context communication – most of the information is either in the physical context or internalized in the person.
- A low-context communication – the mass of the information is vested in the explicit code.

Thinking style (Nisbett *et al.*, 2001):

- Holistic thinking – an orientation to the context or field as a whole.
- Analytic thinking – detachment of the (focal) object from its context.

More recently, with the advancement of technological tools, scholars are able to programmatically scrape eWOM from websites and use sentiment analysis tools to quantify the eWOM message into either a positive or negative valence (for example, Kusawat and Teerakapibal, 2021). Another interesting methodological advancement is the use of topic modeling. This emergent research method allows researchers to explore textual information provided by customers, i.e. online reviews. Research using this technique aims to discover the patterns of how consumers from different cultures focus on different topics in their reviews by directly examining the textual content of the reviews.

## Methodology

Denyer and Tranfield (2009) define the systematic review as a process for reviewing the literature using a comprehensive preplanned strategy to locate existing literature, evaluate the contribution, analyze and synthesize the findings and report the evidence to allow conclusions to be reached about what is known and, also, what is not known. In this paper, we followed Denyer and Tranfield's (2009) guidelines for conducting a systematic literature review in a five-step process consisting of the following:

(1) formulate question(s) for the systematic review;

(2) locate and create an extensive list of prospect relevant research papers;

(3) choose and analyze relevant research papers using predetermined inclusion and exclusion criteria;

(4) analyze and synthesize the relevant literature; and

(5) report the results.

Step 1 has already been addressed in the introduction. Step 5 will be described in the next section. Therefore, this section will focus on addressing Step 2 to Step–4.

*Step 2: locate and create an extensive list of prospect relevant research papers*
According to Verma and Yadav (2021), Scopus and Web of Science are the most reputed bibliometric databases. However, as Scopus is the largest abstract-based bibliometric database, consisting of more than 20,000 peer-reviewed journals from various publishers, its coverage is broader than Web of Science. Thus, we conducted keyword searches for journal articles in the Scopus database.

We limited the sample of empirical studies to those in which both eWOM and cultural dimensions were significant themes. Thus, we used two sets of keywords representing eWOM and cultural differences. The keywords relevant to eWOM were adopted from a previous eWOM literature review. The following are primary and synonym keywords used for eWOM:

"e-WOM," "eWOM," "electronic word of mouth," "online word of mouth," "online customer review," "online consumer review," "online reviewer," "online rating," "online review," and "user generated content."

Below are the primary and synonym keywords representing cultural differences:

"Hofstede," "cultural dimensions," "cultural differences," "cross-culture," "cross-cultural study," "individualism," "collectivism," "power distance," "masculinity," "uncertainty avoidance," "long-term orientation" and "indulgence."

To conduct an extensive and systematic search, for each search, these two sets of keywords were exhaustively combined using "AND" operators, resulting in a total of 120 search terms. Specifically, the following sets of keywords were used: "e-WOM" AND "Hofstede," "e-WOM" AND "cultural dimensions," "e-WOM" AND "cultural differences," . . ., "user generated content" AND "uncertainty avoidance," "user generated content" AND "long-term orientation," "user generated content" AND "indulgence." We used a Boolean operator "OR" in the article title, abstract and keywords tab of the Scopus database to search for the research papers. The initial search resulted in 141 articles. To avoid missing any relevant articles, we cross-validated this search with the Web of Science and SAGE publication databases. The results indicated that SCOPUS had already incorporated all the papers from the Web of Science and SAGE publication databases.

*Step 3: choose and analyze relevant studies according to predetermined inclusion and exclusion criteria*

Inclusion and exclusion criteria were used to delimit the search results to extract only the most relevant articles for the literature review. Journals are believed to represent the highest level of research through which academics and practitioners acquire information and disseminate new findings (Nord and Nord, 1995). Therefore, the search results were based solely on academic journals. Book reviews, conference papers, research notes and editor prefaces were disregarded because of their limited contribution to knowledge development. After the first inclusion and exclusion criteria had been applied, the number of articles was reduced to 124.

The second exclusion criterion of language was then applied to the search results. Only articles published in English were included in the set of articles for further review. After the second inclusion and exclusion criteria of the English language had been applied, the number of results remained the same at 124 research articles.

The final inclusion and exclusion criteria were based on the quality of the paper and relevancy to the cross-cultural eWOM domain. Only papers that had a defined sample and an empirical methodology were included. The abstract and full text of each article were read in detail to assess their suitability.

In total, 61 journal articles from 38 journals, published from 2006 to the beginning of October 2021, were examined. Thus, the recent trends in eWOM and culture research were captured based on studies published over the past 15 years.

*Step 4: analyze and synthesize the relevant literature*
To organize the main findings of each paper, we followed Cheung and Thadani (2012) in adopting a social communication framework. By identifying who says what to whom and with what effect, this framework incorporates four major elements: the communicator (source) is the person who transmits the communication, the stimulus (content) is the message transmitted by the communicator, the receiver (audience) is the individual who responds to the communication and the response (main effect) is the receiver's response to the message.

In our study, we first outline the current knowledge from the literature. Subsequently, we identify knowledge gaps that can potentially be used for future research.

Figure 1 shows an overview of the research methodology.

## Results

*Descriptive results*
Among the 61 articles on cross-cultural eWOM studies, 54% were published in business, management and accounting journals; 15% were published in social sciences journals; 15% were published in computer science journals and the rest of the papers were published in journals incorporating a mixture of disciplines, such as economics, econometrics and finance, arts and humanities.

The number of published papers shows that the *Journal of Business Research* has published the greatest number of articles (5), followed by the *International Journal of Contemporary Hospitality Management* (4), *International Marketing Review* (4), *International Journal of Hospitality Management* (3), *Journal of Global Marketing* (3) and *Tourism Management* (3). Other journals that have contributed to the topic are: *Electronic Commerce Research and Applications*, *Journal of Electronic Commerce Research*, *Journal of Interactive Marketing*, *Journal of Intercultural Communication Research*, *Journal of International Marketing*, *Journal of Internet Commerce* and *Journal of Marketing Communications*. The articles from these journals combined account for 59.02% of the literature on cross-cultural eWOM. The variety of journal disciplines indicates the widespread and growing interest in cross-cultural eWOM research across various academic fields, such as business, marketing, tourism and management information systems.

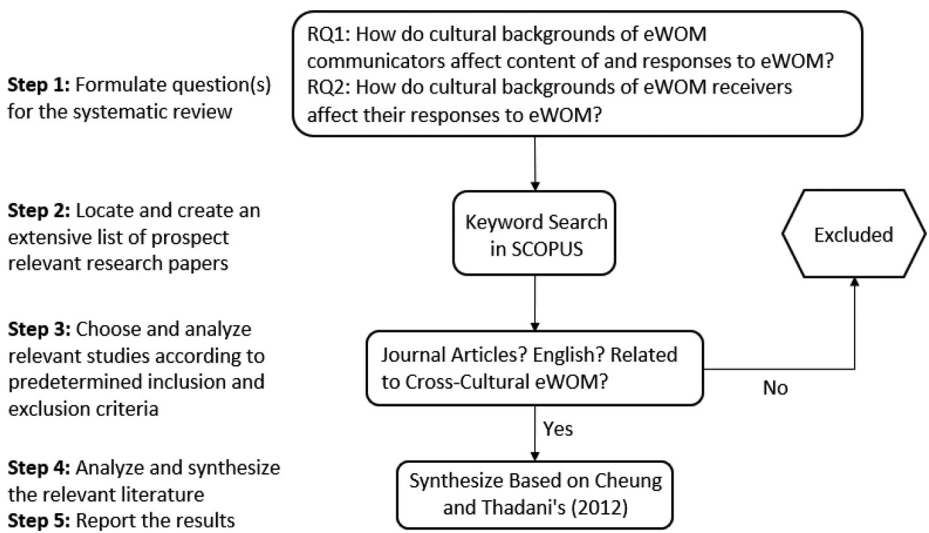

**Figure 1.**
Research
methodology

The earliest article found was published in 2006, which is of no surprise since the major online review websites were launched during the early 2000s. For ease of interpretation, the timeframe was divided into three equal intervals. During the first subperiod (between 2006 and 2011), only four articles (6.56%) were published. In the second subperiod (between 2012 and 2016), the discipline's development accelerated slowly, with 11 research articles (18.03%) published. Since 2017, the growth of cross-cultural eWOM research publications has been phenomenal, with 46 articles (75.41%) having been published by the beginning of October 2021. Overall, the findings show an upward trend starting around the early 2010s in the number of publications on cross-cultural eWOM studies, with a relatively steeper increase toward 2020. This rapid growth depicts the increasingly strong research attention on cross-cultural eWOM.

Among the 61 cross-cultural eWOM studies, Hofstede's cultural dimensions theory was the most used theoretical foundation. Only 15 of the 61 studies based their research on other cultural theories.

*Synthesized cross-cultural eWOM findings*
In this study, we reviewed the 61 cross-cultural eWOM studies and identified variables related to the four key elements (communicators, stimuli, receivers and responses) of social communication. Extending the framework proposed by Cheung and Thadani (2012), Figure 2 depicts the interrelationships among the key elements. Factors associated with each element of the framework are as follows:

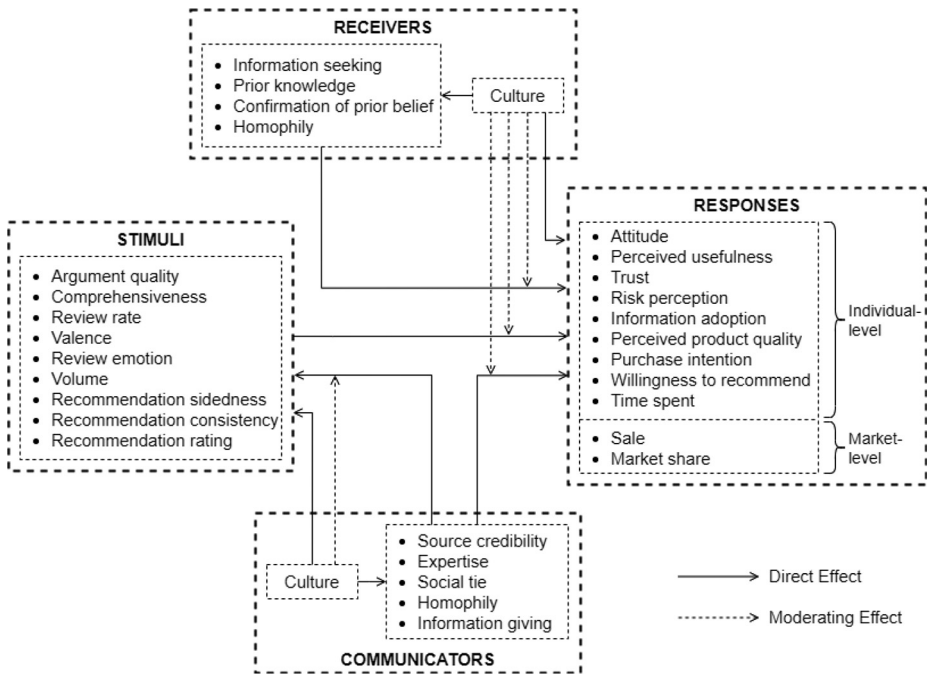

**Figure 2.**
A framework of the cross-cultural eWOM

Communicators:

- Source credibility – message source's perceived ability (expertise) or motivation to provide accurate information.
- Social tie – the extent to which two individuals are socially related.
- Homophily – the degree to which individuals are similar in age, gender, education and social status.
- Information giving – the tendency for an individual to give an opinion about products or services.

Stimuli:

- Argument quality – the persuasive strength of arguments embedded in an informational message.
- Comprehensiveness – the completeness of a message.
- Review rate – the rating given by communicators on a product.
- Valence – whether the message is positive or negative.
- Review emotion – the extent to which communicators express their emotions in the message.
- Volume – number of reviews.
- Recommendation sidedness – two-sided information contains both positive and negative statements, one-sided information contains only either positive or negative statements.
- Recommendation consistency – the degree to which a particular eWOM recommendation is consistent with existing recommendations.
- Recommendation rating – the overall rating given by other readers on an eWOM recommendation.

Receivers:

- Information seeking – the tendency that an individual seeks information about products or services from others.
- Prior knowledge – prior knowledge of the review topic and the platform.
- Confirmation of prior belief – the level of confirmation/disconfirmation between the received information and their prior beliefs about the products or services.
- Homophily - the degree to which individuals are similar in age, gender, education, and social status.

Responses:

- Attitude – reviewer's overall evaluation of a person, messages, products or services.
- Perceived usefulness – the perception of an eWOM message being useful.
- Trust - the perception of the degree of truthfulness of the message.
- Risk perception – the uncertainty a consumer has when making purchase-related decisions.
- Information adoption – a process in which people purposefully engage in using information.

- Perceived product quality – perceived level of product quality.
- Purchase intention – the willingness to purchase a product in the future.
- Willingness to recommend – the willingness to recommend products in the future.
- Time spent – the amount of time spent on searching and considering product choice.
- Sale – the number of products sold.
- Market share – the portion of a market dominated by a product.

Tables 1 and 2 summarize the direct effects of cultural values on these factors. It is worth nothing that Tables 1 and 2 only include relationships with a theoretical base, specifically, those relationships hypothesized by the authors. This section first outlines the current knowledge from the literature in each element. Subsequently, knowledge gaps that can potentially be used for future research are identified.

*Communicators*
Most research studies were in agreement that individualistic consumers engage in more information giving than collectivistic consumers (Fong and Burton, 2008; Cheong and Mohammed-Baksh, 2020; Kitirattarkarn *et al.*, 2021) because individualistic societies appreciate the expression of different opinions, whereas collectivistic cultures focus on the preservation of harmony, respect for hierarchy and the saving of face within the group. Therefore, collectivistic cultures tend to express opinions less to avoid challenging other opinions within the group. These findings were consistent for both horizontal and vertical dimensions of individualism (Choi and Kim, 2019). However, the results were inconsistent

| Variables | PDI + | PDI − | IDV + | IDV − | MAS + | MAS − | UAI + | UAI − | LTO + | LTO − | IND + | IND − |
|---|---|---|---|---|---|---|---|---|---|---|---|---|
| Communicators | | | | | | | | | | | | |
|   Expertise | | | | | | | | 1 | | | | |
|   Homophily | | | | 1 | | | | | | | | |
|   Information giving | | | 3 | | | | | | | | | |
| Stimuli | | | | | | | | | | | | |
|   Comprehensiveness | 1 | 1 | 1 | 1 | 1 | | 1 | 1 | 1 | | 1 | |
|   Review rate | 1 | 3 | 1 | 6 | | 2 | | 4 | 1 | | 2 | |
|   Valence | 1 | | 1 | 2 | | | | 1 | | | 1 | |
|   Review emotion | | | | | | | 1 | | | | | |
|   Volume | | 1 | 2 | | | | | | | | 1 | |
|   Recommendation sidedness | | | 1 | | | | | | | | | |
|   Recommendation consistency | | 2 | 1 | 4 | 1 | | 1 | 1 | 1 | | 2 | |
|   Recommendation rating | 1 | | 1 | | | | 2 | 1 | | | 1 | |
| Receivers | | | | | | | | | | | | |
|   Information seeking | | | | 2 | | | | | | | | |
| Responses | | | | | | | | | | | | |
|   Perceived usefulness | | | | 1 | | | | | | | | |
|   Trust | | | | | | | | 2 | | | | |
|   Purchase intention | | | | | | | | | | | 1 | |

**Notes:** PDI: power distance, IDV: individualism, MAS: masculinity, UAI: uncertainty avoidance, LTO: long-term orientation, IND: indulgence

**Table 1.**
Number of studies with positive/negative effects of Hofstede's cultural dimensions on eWOM factors for the horizontal and vertical dimensions of collectivism (Chu and Choi, 2011; Lee *et al.*, 2018; Choi and Kim, 2019).

Consumers tend to share information with their closely related peers. As collectivism is positively associated with the degree of homophily and social ties (Chu and Choi, 2011; Pezzuti and Leonhardt, 2021), the positive effect of social ties on information giving is more salient in collectivistic cultures (Chu *et al.*, 2020). In addition, Litvin (2019) found that low uncertainty avoidance travelers plan their travels more extensively; therefore, their reviews tend to appear to be from the perspective of high expertise reviewers.

Though the empirical evidence in the literature offered fruitful insights, a few research questions remain open for future research. First, most existing research suggests that collectivistic consumers tend to post less compared with individualistic consumers because consumers from collectivistic cultures want to avoid challenging other opinions within the group. However, none of these studies measured whether this conjecture truly explains this phenomenon. Future research may address this gap by comparing the number of reviews by collectivistic consumers between products with strong conflicting reviews and products with strong congruency among reviews. If the said mechanism is correct, we expect that the number of reviews from collectivistic consumers would be significantly higher in products with strong congruency among reviews. Specifically, future research may examine the interaction between collectivism and congruency among reviews. The main negative effect of collectivism and the number of reviews must be significant, and this negative effect should be attenuated for products with high review congruency. Further, it could even become positive because collectivistic consumers may prefer sharing reviews that are consistent with other reviews to connect with their in-group peers (Lee *et al.*, 2019).

Second, self-enhancement was found to be one of the factors determining eWOM sending behavior (De Angelis *et al.*, 2012). Despite existing arguments on the possibility that the presence of self-enhancement is much weaker in collectivist-dominant Asian culture (Kitayama *et al.*, 1997), no research has yet examined the cross-cultural effect of self-enhancement on the generation of eWOM. Future research may attempt to address this gap.

| Variables | CD + | CD − | GLOBE's IDV + | GLOBE's IDV − | HCC + | HCC − | ATS + | ATS − | HVIC HI + | HVIC HI − | HVIC VI + | HVIC VI − | HVIC HC + | HVIC HC − | HVIC VC + | HVIC VC − |
|---|---|---|---|---|---|---|---|---|---|---|---|---|---|---|---|---|
| **Communicators** | | | | | | | | | | | | | | | | |
| Social tie | | | | | | | | | | | | | 1 | | | |
| Information giving | | | | | | | | | 1 | | 1 | | 2 | 1 | 1 | |
| **Stimuli** | | | | | | | | | | | | | | | | |
| Comprehensiveness | | | | | | | 1 | | | | | | | | | |
| Review rate | 4 | | | | | | | | | | | | | | | |
| Review emotion | | | 1 | | | | | | | | | | | | | |
| Recommendation consistency | 1 | | | 1 | | | | 1 | | | | | | | | |
| **Receivers** | | | | | | | | | | | | | | | | |
| Information seeking | | | | | | | | | | | 1 | | 2 | | 2 | |
| **Responses** | | | | | | | | | | | | | | | | |
| Perceived usefulness | 1 | | | | | | | | | | | | | | | |

**Notes:** CD: cultural distance, HCC: high cultural context (vs low), ATS: analytical thinking style (vs Holistic)

Table 2.
Number of studies with positive/negative effects of cultural distance, GLOBE scores, cultural context, thinking style and HVIC values on eWOM factors

Third, self-efficacy was found to be an antecedent of eWOM (Huang *et al.*, 2009). It would be interesting to examine this relationship across cultures because self-efficacy appears to differ among cultures (Oettingen, 1995).

Finally, previous research suggests that individuation (the willingness to stand out or be different from others) positively affects eWOM forwarding (Ho and Dempsey, 2010). However, as consumers from collectivistic cultures value interpersonal harmony above personal preference, they are less willing to stand out from the group. Therefore, we suspect that the impact of individuation on eWOM forwarding is attenuated in consumers from collectivistic cultures. Future research may attempt to prove this conjecture.

*Stimulus*

Most studies focused on observing the impact of culture on the review rate. Though some studies documented opposing findings (Kim *et al.*, 2018; Stamolampros *et al.*, 2019), the majority found that review rate was related to collectivism (Chiu *et al.*, 2019; Mariani *et al.*, 2019; Mariani and Predvoditeleva, 2019; Stamolampros *et al.*, 2019) and low power distance (Gao *et al.*, 2018; Mariani *et al.*, 2019; Mariani and Predvoditeleva, 2019). This is because customers from countries with a high score on individualism and/or power distance dimensions have higher expectations. Therefore, they have a greater tendency to complain about disconfirmations in the perceived service quality. On the other hand, Stamolampros *et al.* (2019) argued that because individuals from high power distance cultures accept differences in social status, they are more likely to tolerate failures from the more powerful service providers. This argument is in fact very similar to that of studies proposing power distance to be negatively related to review rate. The difference is whether they see the service providers as holding a higher status or lower status. Further, Kim *et al.* (2018) posited that individualistic cultures prioritize enjoyment of life and fun, resulting in a positive attitude and optimistic outlook, which eventually leads to a more positive review rate.

Masculinity is found to be negatively associated with the review rate (Mariani and Predvoditeleva, 2019; Stamolampros *et al.*, 2019). This is because masculine consumers are less tolerant of service failures and perceive themselves to have the power to confront service providers over the unsatisfactory experience (Torres *et al.*, 2014). Additionally, long-term orientated consumers were found to be less likely to give a negative review rate (Mariani *et al.*, 2019) because they are not willing to take the risk of compromising their long-term relationships with the service provider. The relationship between indulgence and review rate was found to be positive (Mariani *et al.*, 2019; Stamolampros *et al.*, 2019) because individuals from indulgent societies have a more positive attitude and are therefore more optimistic and more likely to remember positive emotions. However, individuals from restrained societies are less happy, less likely to remember positive emotions and more pessimistic. Further, uncertainty avoidance was found to negatively affect the review rate (Litvin, 2019; Mariani *et al.*, 2019; Mariani and Predvoditeleva, 2019; Stamolampros *et al.*, 2019). This is because customers with higher uncertainty avoidance are more risk-averse; therefore, they tend to search for the attributes of the product and service, leading to higher expectations. The cultural distance was also found to be related to a negative review rate (Mattson, 2017; Stamolampros *et al.*, 2019). Because when the cultural distance between customers and sellers is higher, there is a higher chance of cultural gaps and differences, and these differences may translate into consumers' dissatisfaction. Similar to the review rate, review valence was found to be associated with indulgence (Wen *et al.*, 2018), high power distance (Wen *et al.*, 2018) and low uncertainty avoidance (Fang *et al.*, 2013), with mixed findings in individualism dimension.

In this review, it was noted that several studies also paid much attention to volume and recommendation consistency. The findings showed that the volume of eWOM messages was significantly more in individualistic cultures (Fang *et al.*, 2013; Chiu *et al.*, 2019). This is in line with the previous notion that individualistic consumers tend to engage in information giving more than collectivistic consumers. Also, higher volume was found in low power distance societies than in high distance power societies because individuals from high power distance are recognized as introverts and more reluctant to express their feelings (Fang *et al.*, 2013). With online movie review data, Chiu *et al.* (2019) found that indulgence positively influences eWOM volume.

For recommendation consistency, various cultural theories were applied across the studies reviewed to explain the difference between East Asians and Westerners. Using Hofstede's theory, Chiu *et al.* (2019) posited that members of individualistic Western societies are more likely to value freedom of expression, while members of collectivistic East Asian societies are more likely to seek group consensus rather than directly expressing their opinions. On the other hand, Wang *et al.* (2021) proposed that because consumers from individualistic cultures are more likely to engage in voice behaviors and are more involved, they are more susceptible to the anchoring effect of prior eWOM. Drawing on thinking style theory, Kim *et al.* (2018) posited that East Asians (holistic thinking style) are more likely to be influenced by their surroundings and peripheral cues, such as prior eWOM, while Western customers (analytical thinking style) tend to see an object as independent of its context and are therefore likely to evaluate a service or product without external influences. These studies are inconsistent both in their findings as well as theoretical background. Future research may attempt to clarify these inconsistencies.

### Receivers

The findings of the studies included in this review were mostly consistent with the notion that collectivistic consumers are more likely to seek eWOM than individualistic consumers (Fong and Burton, 2008; Cheong and Mohammed-Baksh, 2020). This is because collectivistic cultures perceive that the act of using the information gained from a referent group enhances their relationships. On the other hand, individualistic societies are independent and rely less on others when they seek information. Other research delved deeper into the HVIC. The results showed that both horizontal (Chu and Choi, 2011; Lee *et al.*, 2018) and vertical (Lee *et al.*, 2018; Choi and Kim, 2019) collectivism were positively related to information-seeking behavior. However, Choi and Kim (2019) argued that some individualistic consumers also seek information from eWOM. Specifically, vertical individualistic consumers seek eWOM information because they believe that winning or doing better than others is crucial; therefore, they use all information available to them to be more competent.

### Responses

Based on our framework, the response is a function of stimuli, communicators and receivers, and these relationships are moderated by the cultural backgrounds of the receivers of eWOM. Tables 1 and 2 shows that only three factors are directly affected by the cultural values of the receivers: perceived usefulness, trust and purchase intention.

Applying Hofstede's cultural theory, Noh *et al.* (2013) found significant effects of collectivism on perceived usefulness. Further, Kim *et al.* (2018) found that cultural distance negatively affects perceived usefulness. This is because consumers who read reviews written by reviewers from similar cultural backgrounds assume that the reviewers are also similar in terms of preferences and attitudes, leading to an increased perceived usefulness of reviews. In terms of the trust, because uncertainty avoidance is related to the risk averseness of individuals, people from high uncertainty avoidance cultures tend to be more sensitive to information from unknown sources, which is generally the case with eWOM. In line with this argument, Furner *et al.* (2014) documented that consumers with high uncertainty avoidance exhibit lower trust. Regarding purchasing intention, compared with consumers from restraint societies, consumers from indulgent societies are usually more optimistic, more positive and happier; therefore, they have a higher tendency to engage in impulsive buying behaviors, making more spontaneous purchase decisions regarding products with positive reviews (Ruiz-Equihua *et al.*, 2020).

Despite the various contributions offered in these studies, our synthesis suggests that additional research efforts may be undertaken to further develop the cross-cultural eWOM literature. Though the extant literature has investigated the moderating roles of culture on various relationships, it is fragmented, with one study for each relationship. Future research should attempt to conduct a replicated study for these relationships.

Second, research findings showed that eWOM significantly affects levels of trust and loyalty (Awad and Ragowsky, 2008). However, the extant literature suggests that there exists a cross-cultural difference in customer loyalty (De Silva Kanakaratne *et al.*, 2020). Therefore, future research may attempt to discover differences in the relationship between eWOM and consumer loyalty across cultures. The findings may assist practitioners in leveraging their target customers' cultural orientation when designing their loyalty reward programs.

Another venue of research concerns the anonymous nature of eWOM. This characteristic results in an important aspect unique to eWOM, fake online reviews, and this aspect may actually be its own biggest drawback. It is well known that product sellers manipulate product reviews to increase sales (Chen *et al.*, 2011). In different regions, the propagators possibly adopt different strategies to disseminate fake reviews. Thus, future research may explore cultural differences in the posting of and responses to fake reviews.

Fourth, research suggests that consumers seek out both the most negative reviews and the most positive reviews to get a range of customer feedback. However, negative reviews are given more importance by consumers (Park and Lee, 2009). Future research may examine this negativity bias across cultures. It is likely that different cultures are predisposed to a different spectrum of eWOM valence.

Finally, visual eWOM is an emerging area of research that has received no attention in the cross-cultural context. Visual eWOM can be either a product review or the increasingly popular product "unboxing" videos posted on Instagram, YouTube, TikTok and other social media platforms. Using information processing theory, future research should investigate how consumers decide which visual eWOM to trust and which visual eWOM to reject.

## Conclusion

As online platforms become increasingly globalized, scholars and practitioners are paying increasing attention to the topic of cross-cultural eWOM. Thus, it is crucial to examine the cross-cultural eWOM literature to understand its current state. This study reviewed the 61 cross-cultural eWOM studies and identified variables related to the four key elements (communicators, stimuli, receivers and responses) of social communication. Our results indicated that eWOM indeed differs among different cultures. The source of cultural differences can be either the communicators or the receivers. This review found that the cultural background of a communicator only affects the characteristics of their eWOM message (stimuli). Similarly, the cultural background of a receiver only affects their response to an eWOM message. Further, the cultural background of the receiver moderates the effects eWOM characteristics and communicators have on their response. This review also found that most of the 61 studies were built upon Hofstede's cultural dimensions theory. This is consistent with Engelen and Brettel's (2011) review on cross-cultural theory in marketing. As pointed out by Nakata (2009), scholars from different fields tend to present their findings at different conferences and publish them in different journals. This lack of interaction with other fields may cause cross-cultural scholars to only build upon popular theory (i.e. Hofstede's framework) without considering alternatives in related fields. As the adoption of other cultural frameworks is rare in eWOM studies, future research should attempt to use alternative cultural frameworks in their research to avoid this situation.

This study theoretically contributes to the literature in two ways. First, this study provides a comprehensive overview of the current status of knowledge within the domain of cross-cultural eWOM. To our knowledge, this is the first study to systematically review cross-cultural eWOM research and provide a framework. The provided research framework can serve as an important foundation for future research, as it integrates key elements of cross-cultural eWOM. Second, our findings suggest that research on cross-cultural eWOM is very fragmented. Replicated studies on the same constructs are rarely conducted; this is especially the case for variables in the response element. This is a problem because consumer behavior is one of the most important elements marketers pay attention to. Without replication, the research community and practitioners may find it hard to trust and act upon a study's results. We strongly believe that future research could use our framework as a basis to empirically confirm existing findings or investigate areas where the findings are scarce and/or conflicting. For practitioners, our results illustrate how consumers' cultural backgrounds lead to differences in the generation of and reaction to eWOM. Practitioners can use our results as a guideline to determine which factors could drive the success of their marketing outcomes. For instance, most studies suggested that individualistic consumers tend to write negative reviews. Therefore, for campaigns that allow customers to test the product, international marketers may allocate the majority of their budget to geographic regions with collectivistic customers and less to geographic regions with individualistic customers.

This study is not without limitations. Because we conducted a structured database search using SCOPUS and included only journal articles, there is a certain possibility that not all relevant articles were identified because they are not indexed in SCOPUS or they are conference papers. Future research may attempt to be more inclusive in their database search. In addition, there are a limited number of cross-cultural eWOM studies; therefore, it is not yet feasible to conduct a meta-analysis. Because a meta-analysis statistically combines the results of multiple studies, it is able to draw a more quantitative conclusion on the relationships of constructs. As more research on cross-cultural eWOM emerge, future research may attempt to conduct a literature review with this methodology. Despite these limitations, we hope that our findings are able to assist interested parties in grasping the current state of cross-cultural eWOM literature.

**Corresponding author**

Surat Teerakapibal can be contacted at: suratt7@tbs.tu.ac.th